 \documentclass[9pt,prd,aps,amssymb, 
 amsmath,tightenlines]{revtex4}
 \usepackage{graphics}
 \usepackage[pdftex]{graphicx}

 \usepackage{amsmath,amssymb,amsthm}

\begin{document}


\title{
 Carath\'eodory, Finite Resources  and the Geometry of Arbitrage
}

\author{ B. K. Meister}

\email{bernhard.k.meister@gmail.com}

\date{\today }

\begin{abstract}
\noindent
Carath\'eodory's axiom of adiabatic inaccessibility states that, in any neighborhood of a thermodynamic state, certain states remain unreachable via adiabatic processes. Non-arbitrage mirrors this topological restriction in finance. 
Preserving this constraint in resource-limited systems identifies the exponential family not as a modeling convenience but as the requisite geometric structure unifying both domains. \end{abstract}
   \noindent

\maketitle

\section{Introduction}
\label{sec:introduction}
 \noindent
 Carathéodory's second law begins with a seemingly local statement -- in any neighborhood of an equilibrium state, there exist states that are adiabatically inaccessible\footnote{A more precise formulation of  Axiom II: `In jeder beliebigen Umgebung eines willk\"urlich vorgeschriebenen Anfangszustandes gibt es Zust\"ande, die durch adiabatische  Zust\"ands\"anderungen nicht beliebig approximiert werden k\"onnen'.} -- yet it is in effect a global constraint. 
 Carathéodory\cite{caratheodory1909} showed that, for the `simple systems' he considered, local integrability follows, and from this, a global foliation into entropy surfaces\footnote{Buchdahl \cite{buchdahl1966, buchdahl1975} offers a minimalist restatement of classical thermodynamics from conceptual foundations.}. But a global extension from local integrability for arbitrary systems is not automatic, as Boyling's counterexample\cite{boyling1968} demonstrates\footnote{Buchdahl \& Greve\cite{buchdahl} and others had earlier noted this limitation.}.
 
   \noindent
 This gap is not  a thermodynamic curiosity.  No-arbitrage in finance presents a direct analogue.  Within any  small neighborhood of a portfolio state, local non-arbitrage ensures the existence of a local pricing function. This local consistency does not guarantee global market stability. A trader might execute locally fair trades along a closed loop for net gain. Without additional structure, the local foliation that guarantees path-independence in the small fails to extend globally.

\noindent
To close this gap,  an operational constraint\footnote{
This aligns with the ``ultra-finitism" of A. S. Esenin-Volpin\cite{esenin1970}, who rejected the infinite-precision continuum as a mathematical fiction, arguing that physical reality is bounded by the feasibility of step-wise construction. If a value cannot be constructed in a finite number of observable steps,  it does not exist.
  Esenin-Volpin went further, questioning the existence of even very large finite numbers.} 
 is imposed: the market's probabilistic state must remain describable with finite resources regardless of scale -- ergo, information about the market state must be compressible into a fixed set of summary statistics whose number does not grow with the size of the system.

\noindent
The Pitman-Koopman-Darmois theorem (PKD) characterizes the unique consequence of this condition for independent and identically distributed observations: if a family of distributions admits a sufficient statistic whose dimension does not grow with sample size, then it must be an exponential family\footnote{In the Information Geometry of Amari, an 'exponential family' is intrinsically defined by a finite set of sufficient statistics (making 'finite-rank' technically redundant).}. 
Finite-resource description thus mandates the exponential family for market observables.

\noindent
This result depends crucially on the IID assumption\footnote{The IID assumption is the simplest setting in which the thermodynamic limit is well-defined and the PKD theorem applies.  While real markets may exhibit dependencies and non-stationarities, this idealized regime defines the baseline where our sufficient condition holds; violations signal a possible departure from the simple equilibrium described by our framework, though more general equilibrium concepts (such as those of Lieb-Yngvason) may still apply.}. Consequently, our framework excludes regimes where this assumption fails.  In thermodynamics, this would be small systems where fluctuations dominate (the `Resolution Gap') and critical points where correlation lengths diverge (the `Correlation Boojum'). In finance this includes for example  periods of market instability, liquidity crises, and structural breaks where `market temperature' becomes ill-defined due to non-stationarity.

\noindent
Under the operational constraint  imposed, the exponential family is thus not a modeling convenience but the necessary geometric structure that preserves coherence of local no-arbitrage conditions across scales.

  \noindent

 \noindent
 Our choice forgoes the axiomatic generality of approaches like Lieb \& Yngvason~\cite{LY} in favor of a more constrained and operational structure. While their formulation establishes entropy under general conditions, it places no bound on descriptive complexity. Our operational approach aligns with viewing thermodynamic and financial problems through the lens of information. For systems where finite-resource describability holds -- a class that includes many thermodynamic settings and equilibrium states of observable markets -- the distribution belongs to the exponential family, yielding a globally convex potential with a well-behaved Fisher metric. This provides a sufficient condition for thermodynamic-like behavior in finance\footnote{Buchdahl in his publications on thermodynamics  lets ``physical intuition take precedence over mathematical subtleties''. We share his stance, even if it chafes against the realities of finance and statistics.}.

 \noindent
 Two characters from Lewis Carroll frame the challenge. The   Snark represents the elusive goal of a globally integrable market model. The Boojum is where this structure `softly and suddenly vanishes away': in the `resolution gap' of small systems, the diverging correlations at critical points, and during the `regime shifts' of non-stationary markets -- where the foliation tears, description length becomes infinite, or 'market temperature' loses meaning.

\noindent
Section II reviews the thermodynamic laws and their financial analogues. Section III presents Boyling's counterexample in detail. Section IV introduces the PKD theorem and shows how finite-resource describability leads to the exponential family.  Section V extends the geometry to dynamics via Onsager reciprocity. Section VI concludes.

\section{Thermodynamic Laws: Prerequisites and Constraint}
\label{sec:thermo_blueprint}
\noindent
Market structure, by thermodynamic analogy, comprises three  layers.
%
 The first two are prerequisites that make the third meaningful; only the third imposes the geometric constraint.

\subsection{Zeroth Law: State Space Exists}
\label{subsec:zeroth}
\noindent
The Zeroth Law establishes that thermal equilibrium is transitive: if system $A$ equilibrates with $B$ and $B$ with $C$, then $A$ equilibrates with $C$. 
This partitions state space into equivalence classes -- surfaces of constant temperature -- yielding a foliated manifold for thermodynamics.
In finance, the Law of One Price plays the analogous role.


\subsection{First Law: Path-Independence}
\label{subsec:first}
\noindent
The First Law asserts path-independence of work in insulated systems: $\delta W_{\text{ad}} = -dU$ establishes internal energy $U$ as a state function. This defines what counts as an `adiabatic process' (one with $\delta Q = 0$) and foliates state space into isoenergetic surfaces.
In finance, this corresponds to the self-financing condition. Namely, changes in portfolio value are given solely by price changes of the constituent assets, with no external inflows or outflows -- the analogue of an adiabatic process. 

\subsection{Second Law: Geometric Constraint} 
\label{subsec:second}
\noindent
Carath\'eodory's axiom states that in any neighborhood of an equilibrium state, there exist states inaccessible via adiabatic processes. 
In finance, the non-arbitrage axiom states that in any neighborhood of an allowed market state, some markets states are inaccessible via non-arbitrage processes.

\section{The Boyling Counterexample: A Toy Model of Market Incoherence  } 
 
\noindent
 Building on the Zeroth and First Law analogues from Section II -- the Law of One Price and the self-financing condition -- the wealth change from a trading strategy is represented by a 1-form $\psi$
on the state space. To understand why local integrability does not guarantee global stability, consider a `toy market' defined on a two-dimensional plane $\mathbb{R}^2$. In this model (directly taken from Boyling \cite{boyling1968}), the infinitesimal flux of wealth is described by the 1-form:
$$\psi = y^3(1 - y)^2 dx + [y^3 - 2(1 - y)^2] dy.$$
This expression satisfies the condition for local consistency.  
In this context, local consistency is synonymous with local non-arbitrage: it means that within any sufficiently small neighborhood, the market is integrable. In such a neighborhood, there exists a relationship $\psi = f dg$, where $g$ is a local potential and $f$ is a local integrating factor (the inverse "market temperature"). 

\noindent
Under a microscope, this relationship ensures the state space appears `layered', preventing any infinitesimal closed-loop trading strategy from extracting risk-free gain.  Because the local flux is tied to a local potential, any round-trip trade within a small patch must return to its starting value. 

\noindent
This local non-arbitrage does not generalize to the system as a whole.  While the manifold is smooth and simply connected, the field $\psi$ is fundamentally twisted, as 
the local relationship $\psi = f dG$ fails to cohere into a single, global potential. On  the strip $x\in R$ and $0<y<1$, $\,\,\, G(x,y)$ has the form $$ h\left(x + \frac{1}{y^2} + \frac{1}{1-y}\right).$$ 
The  integrating factor $f$ must satisfy on the strip $$f(x, y) h' \left(x + \frac{1}{y^2} + \frac{1}{1-y}\right) = y^3(1 - y)^2.$$
The contradiction is then found in the asymptotic requirements of $h'(t)$. For the market to remain globally consistent, $f$ must be continuous and non-zero across the entire domain, requiring it to tend toward finite, non-zero limits as $y \to 0$ and $y \to 1$. As a consequence, $h'(t)$ is forced to satisfy two mutually exclusive decay rates for $t \to +\infty$: it must scale as $t^{-3/2}$ to satisfy the $y \to 0$ boundary and as $t^{-2}$ for the $y \to 1$ boundary,
leading to a contradiction.

\noindent
The geometry is thus inherently twisted, representing a geometric manifestation of an infinite-rank system. This occurs because the local integrating factor (temperature) cannot be scaled across the manifold without the description length of the market state exploding.

\noindent
The missing link is this: unbounded decay rates prevent the local potentials from cohering into a single global function. As is shown in the next section, imposing finite-resource describability eliminates this obstruction.

\noindent
Lack of a single global function enables  general arbitrage over long cycles; a trader can execute a series of locally `fair' trades and return to the starting point with a net gain. Without the exponential family identified by the PKD theorem or something analogous, local consistency cannot prevent a 'perpetual motion machine' as market complexity scales. The gap can be closed by moving beyond the local topology to the information-theoretic requirement of a fixed-dimension sufficient statistic.

\noindent
The Boyling counterexample thus reveals a hierarchy\footnote{ Instantaneous Non-Arbitrage (Short-Path Consistency): Corresponds to local integrability ($\psi \wedge d\psi = 0$). This ensures that for any immediate portfolio adjustment, a local "up" and "down" direction exists. It forbids "short-path" wealth extraction but remains blind to global topology. \\
  General Arbitrage (Round-Trip Incoherence): A global condition requiring path-independence. In a system lacking global integrability, a trader can execute a "long-path" cycle—a series of trades that return to the same state-space coordinates but result in a net gain in wealth.}: a market can satisfy instantaneous non-arbitrage (local integrability) while permitting general arbitrage (global incoherence). Closing this gap requires additional structure.

  \section{The PKD Theorem as the Geometric Stabilizer} 
  \noindent
  The Boyling counterexample established that a market can be locally integrable everywhere yet globally twisted, permitting closed-loop arbitrage. The question is what additional structure can prevent this. The PKD theorem\footnote{The theorem's regularity conditions (smoothness) and the requirement of constant support are assumed to be satisfied for the families of distributions considered here.} shows that, under an operational constraint on describability, the distribution must be exponential family.
  
  \noindent
A conditional distribution belongs to the exponential family if it can be written in the form:
$$p(X|\theta) = \exp\left( \langle \theta, T(X) \rangle - \psi(\theta) \right),$$
where $X$ represents the observable variables (prices, returns, volumes), $\theta \in \mathbb{R}^d$ is the latent market state (e.g., risk factor exposures), $T(X)$ is a vector of sufficient statistics that capture all information in $X$ relevant to $\theta$, and $\psi(\theta)$ is the log-partition function that ensures normalization. The sufficient statistics $T(X)$ serve as the macroscopic observables -- the quantities that summarize the market's microscopic activity.

\noindent
The exponential family is not an arbitrary choice: its log-partition function $\psi(\theta)$ is globally convex. Its Hessian, the Fisher information metric $g_{ij} = \partial_i \partial_j \psi$, is therefore positive-definite everywhere. 

\noindent
The Boyling construction reveals incompatible scaling requirements for the integrating factor as one moves across the manifold. In statistical terms, this incompatibility means that describing the market state in different regions would require different sufficient statistics -- or a sufficient statistic of growing dimension -- to capture the local structure. The description length, therefore, cannot remain fixed.   

  \noindent
  To see why this forces an infinite-dimensional sufficient statistic, suppose contrarywise that a fixed-dimensional statistic $T(X) \in \mathbb{R}^d$ sufficed for all regions. Then the conditional distributions $p(X \mid \theta)$ would belong to a $d$-dimensional exponential family, with potential $\psi(\theta)$ globally convex. The Fisher metric $g_{ij} = \partial_i\partial_j\psi$ would then be positive-definite everywhere. But Boyling's construction shows that any such global metric would have to satisfy two incompatible scaling regimes at the boundaries -- a contradiction. Hence, no fixed-dimensional sufficient statistic can exist within the IID regime that defines the scope of our analysis\footnote{  One might ask: could a fixed-dimensional statistic exist without the family being exponential? The PKD theorem proves this is impossible under IID. Hence the chain -- fixed-dimensional statistic $\Rightarrow$ exponential family $\Rightarrow$ convex $\psi$ $\Rightarrow$ positive-definite metric -- is necessary.}. The Boyling twist therefore precludes any finite-dimensional sufficient statistic.

  \noindent
  The functional contradiction in Boyling's decay rates ($t^{-3/2}$\, vs. \,$t^{-2}$) is not merely a topological curiosity; it is the geometric manifestation of statistical non-sufficiency. In information-geometric terms, the inability to find a single, globally valid integrating factor implies that the system possesses infinite rank. 

  \noindent

\noindent
Applying this to the Boyling setting, the contradictory asymptotic regimes correspond precisely to a violation of the fixed-dimension condition. The manifold contains regions that, if they were to support a probabilistic interpretation, would require different sufficient statistics -- or a sufficient statistic of growing dimension -- to capture their structure. The Boyling twist is thus a geometric signature of a system that fails the finite-resource describability constraint.

\noindent
For an exponential family, the potential $\psi(\theta)$ is globally convex, ensuring the Fisher metric remains  invertible across the entire manifold. This is precisely what Boyling's construction lacks, and it is exactly what finite-resource describability, via PKD, guarantees: a fixed-dimensional sufficient statistic requires the exponential family\footnote{While infinite-rank integrable forms may exist as mathematical possibilities, they fail the resource-constraint of a stable, macroscopic "temperature" by requiring an unbounded description length.}. This positive-definiteness forbids  global arbitrage\footnote{This establishes global coherence for the statistical manifold and for the wealth potential defined by the self-financing condition. Other financial quantities, if they are smooth functions of the state, retain this property. However, it is possible that for example securities with non-smooth payoffs  may require additional analysis. This is left as an open question.}.

 \subsection{Thermometer Failure} 
 \noindent
 The previous argument shows that finite-resource describability forces the market into the exponential family. This is   equilibrium geometry. A `thermometer test' can now be formulated. When does this geometry not describe observable markets? 
In at least three regimes: for small 
$N$ where fluctuations dominate (Resolution Gap); at critical points where correlations diverge (Correlation Boojum); and for non-stationary markets where the IID precondition fails (Regime Shifts). 

  \noindent
In a discrete setting -- viewing the market as a Directed Acyclic Graph of transactions -- establishing equilibrium conditions is simpler: local exchange rates around any elementary loop (e.g., triangular arbitrage) must sum to zero in log-space\footnote{Because exchange rates are multiplicative, no-arbitrage across a cycle requires the product of rates to equal unity ($R_{12} \cdot R_{23} \cdot R_{31} = 1$). In log-space, this product becomes a linear summation ($\sum \log R_{ij} = 0$). }.
 As before, local consistency on the graph does not guarantee a global potential if the network topology allows large-scale cycles\footnote{Transitory `cyclic arbitrage' observed in liquid markets is thus interpreted not as a topological Snark, but as an extrinsic information flux (noise plus signal) that momentarily pushes the market off the rigid, dually flat manifold considered in equilibrium finance. Such fluctuations are not internal to the state manifold's geometry but represent a departure from the  equilibrium description.}.

\noindent

\noindent

\noindent

\section{Onsager Reciprocal Relations
 and Market Dynamics}
\label{sec:onsager}
\noindent
The preceding sections established the equilibrium geometry. This section connects this geometry to market dynamics, showing how symmetry of the transport matrix -- first studied by Onsager\cite{onsager1931a, onsager1931b} -- emerges from micro-structural scaling or a gradient-flow hypothesis.  
\subsection{Deriving Detailed Balance from Microstructural Scaling}
\noindent
In the standard continuous-time limit of market dynamics -- common to diffusion models -- price fluctuations are dominated by volatility rather than drift over short horizons. Specifically, fluctuations scale as $\mathcal{O}(\sqrt{\Delta t})$ while directional drift scales as $\mathcal{O}(\Delta t)$\footnote{This separation of scales assumes continuous price paths. Jumps, when present, introduce an additional component: jump events occur with probability $\mathcal{O}(\Delta t)$ and their sizes scale independently of $\Delta t$. At leading order as $\Delta t \to 0$, the continuous component still dominates local dynamics, but jumps can in principle break detailed balance. For markets where jump risk is not hedgeable, the detailed balance condition should be understood as applying to the continuous part of the evolution. We leave the pure-jump case for future work.}. 
This separation justifies detailed balance in the local frame, since within each isentropic leaf, forward and reverse transitions become statistically indistinguishable at leading order $\Delta t \to 0$.

\noindent
In the linear response regime, fluxes respond to thermodynamic forces via a transport matrix $L_{ij}$:
\begin{equation}
    \dot{\eta}_i = \sum_j L_{ij} X^j\nonumber
\end{equation}
The thermodynamic variables are
\begin{itemize}
    \item \textbf{Forces ($X^j$):} These are the gradients of the potential $\partial_j\psi$. They represent the `informational pressure' or `stress' pushing the system away from its current equilibrium $\theta$.
    \item \textbf{Fluxes ($\dot{\eta}_i$):} These are the expectation flows, representing the rate of change in market volume or holdings ($\eta_i$) as the system moves to resolve those pressures.
    \item \textbf{Transport Matrix ($L_{ij}$):} This serves as the kinetic linkage (conductance) that determines the magnitude of flow for a given force.\footnote{By inversion, one can define the Price Impact Matrix $M = L^{-1}$, where $X^i = \sum_j M_{ij} \dot{\eta}_j$. In this dual view, the symmetry $M_{ij} = M_{ji}$ ensures that volume shocks (`fluxes') push prices (`forces'). 
    }
\end{itemize}

\noindent
Detailed balance implies $L_{ij} = L_{ji}$. Asymmetry   would permit net circulation around closed loops in state space -- the dynamical analogue of the Boyling twist -- enabling cyclic wealth extraction. Symmetry thus emerges as a necessary condition for the absence of dynamical arbitrage.

\subsection{Gradient Flow as a Sufficient Condition}
\noindent
A complementary connection  arises if it is assumed that market dynamics follow a gradient flow on the statistical manifold. If the system relaxes according to the natural geometry of the space:
\begin{equation}
    \dot{\eta}_i = \sum_j g_{ij} X^j\nonumber
\end{equation}
where $g_{ij}$ is the Fisher information metric. In this case, the transport matrix $L_{ij}$ is identical to the metric $g_{ij}$. Because the Fisher metric is a Hessian ($g_{ij} = \partial_i\partial_j\psi$), it is inherently symmetric. Consequently, $L_{ij} = L_{ji}$ is satisfied automatically.

\noindent
Under this hypothesis, the dynamics inherit the full geometric structure, 
and the path to equilibrium is  determined by the curvature of the manifold\footnote{When the transport matrix is symmetric, the informational cost of moving between equilibrium states is path-independent and quantified by the Bregman divergence \(D(\eta \| \eta') = \psi(\theta) + \phi(\eta') - \sum_i \theta_i \eta_i'\), where \(\phi\) is the dual potential related to \(\psi\) by Legendre transform. This divergence is generally asymmetric; and if the transport matrix were asymmetric, the work done in a round-trip strategy would not sum to zero, allowing for arbitrage.}.

\section{Conclusion: Holding the Boojum at Bay}\label{sec:conclusion}
\noindent
Carathéodory's theorem -- that adiabatic inaccessibility yields a global entropy --holds only for `simple systems'. As Boyling's counterexample shows, this does not extend to complex systems.

\noindent
This vulnerability is resolved by requiring the market's state to be observable via finite resources. This operational choice, via the PKD theorem, forces the exponential family -- whose globally convex potential $\psi(\theta)$ ensures that local no-arbitrage coheres into a global potential. \noindent

\noindent
Only  limits on resources can reasonably ensure  that the market's foliation does not `softly and suddenly vanish away -- for the Snark was a Boojum', and  global loops do not  extract wealth despite local fairness. 
The absence of `unlimited money pumps'\cite{meister2022,meister2023} accords with finite-resource constraints, and the exponential family provides the sufficient geometry to hold the Boojum at bay. 

\noindent
Open questions include: (1) Can the finite resource condition be formulated beyond the PKD theorem?
(2)  Can the gradient-flow hypothesis in Section V be derived  more simply?
 
\end{document}